\documentclass{aa}

\usepackage{graphicx}
\usepackage{txfonts}

\begin{document}

    \title{Hanle signatures of the coronal magnetic field in the linear polarization of the  hydrogen L$\alpha$ line}

       \author{M. \textsc{Derouich}\inst{1}\thanks{Present address:  Colorado Research Associates Division, NorthWest Research Associates, Inc., 3380 Mitchell Ln., Boulder, CO 80301, U.S.A.}, F. \textsc{Auch\`ere}\inst{1}, J. C. \textsc{Vial\inst{1}}, and   M. \textsc{Zhang}\inst{2}}

    \offprints{M.  \textsc{Derouich}\\e-mail: moncef.derouich@ias.u-psud.fr}
    
   \institute{Institut d'Astrophysique Spatiale, CNRS-Universit\'e Paris-Sud 11, 91405 Orsay Cedex, France\\
         \and
             National Astronomical Observatory, Chinese Academy of Sciences, Beijing 100012, China\\
            }
   \date{Received  06 July 2009 / Accepted    21 October 2009  }
    \titlerunning{Hanle effect in the coronal L$\alpha$    line}
    \authorrunning{\textsc{Derouich} et al.}
\abstract{}{ This paper is dedicated  to the assessment of the validity of future coronal  spectro-polarimetric observations  and to prepare their interpretation in terms of the magnetic field vector.  }{ We assume that the polarization of the  hydrogen coronal   L$\alpha$ line
 is    due  to anisotropic scattering of an incident  chromospheric radiation field. The anisotropy is due to geometrical effects  but also to the inhomogeneities of the chromospheric regions which we model by using Carrington maps of the L$\alpha$. 
Because the corona is optically thin, we fully consider the effects of the integration over the line-of-sight (LOS). As a modeling case, we include a dipolar magnetic topology perturbed by a non-dipolar magnetic structure arising from  a prominence current sheet in the corona. The spatial variation of the hydrogen density and the temperature   is taken into account. We determine the incident radiation field developed on   the tensorial basis  at each point along the LOS. Then, we calculate the local 
emissivity vector  to obtain integrated Stokes parameters with and without coronal magnetic field.}{ We show that the Hanle effect is an interesting    technique for interpreting the scattering polarization of the  L$\alpha$ $\lambda$1216 line in order to diagnose the coronal magnetic field. The difference  between the calculated polarization and the zero magnetic field polarization gives us an estimation of the needed polarimetric sensitivity in  future polarization observations.  We also obtain  useful indications about the optimal observational strategy.}{Quantitative interpretation of the Hanle effect on the scattering linear polarization of L$\alpha$      line  can be a crucial source of information about  the coronal magnetic field at a height over the limb $h$ $ < 0.7 \; R_{\sun}$.  Therefore,  one needs the development   of   spatial instrumentation to observe this line. }


\maketitle

 \section{Introduction}
 One of the most powerful tools for the diagnostics 
of magnetic fields in the Sun is the interpretation of 
polarimetric observations (e.g. the monograph by Landi Degl'Innocenti 
\&  Landolfi 2004 and the recent review by Trujillo Bueno   2009).   However, these diagnostics   are  mostly  concerned with   the fields at the photospheric    and      chromospheric levels. The coronal magnetic field presents more  intrinsic difficulties to measure and interpret. This is especially true for the case of the UV coronal lines. Only rather recently, Raouafi et al. (2002) performed the first measurement and interpretation of  the   linear polarization of a UV line (O {\sc vi} $\lambda$1032 line)  polarized under anisotropic scattering by the underlying solar radiation field. In addition,      Manso Sainz \& Trujillo Bueno (2009) proposed 
a  polarizing mechanism showing the    adequate sensitivity of other coronal UV lines to the direction of the magnetic field.  
 These successful works suggest   that  new UV polarimeters  with high sensitivity associated  with theoretical and numerical modeling obtained with a high degree of realism are a fundamental step to be performed in order to extract information on the coronal plasmas. In this context, the Hanle effect on the L$\alpha$ polarization constitutes an excellent opportunity which merits to be   exploited. 

The scattering polarization of the coronal L$\alpha$ line of neutral hydrogen, 
which we are revisiting in this paper,  has been computed by   
Bommier \& Sahal-Br\'echot (1982) and by Trujillo Bueno et al. (2005).   These   authors, however,  neglected the effects of the integration over    the line-of-sight (LOS) by considering  a local position of the scattering hydrogen atom. Since the corona is optically thin, the LOS integration problem  has to be   solved.   Fineschi et al. (1992) treated  the case of the L$\alpha$ line polarization and   took into account the LOS integration.  However, Fineschi et al. considered   the effect of a  deterministic  magnetic field vector  having a direction and strength  {\it independent of the position of the   scattering volume}. They also treated the case of   a random magnetic field.

To improve upon these previous works,   we take into account the variation of the direction and the strength of the magnetic field   for each scattering event     along the LOS.     The   calculation of the polarization generated by scattering   depends strongly on the level of anisotropy of the incident radiation, which in turn depends strongly  on the geometry  of the scattering process  and the brightness variation of the chromospheric regions. In order to  accurately compute the degree of the anisotropy   at each scattering position, we  use    Carrington maps of the chromospheric incident radiation of the  L$\alpha$ line obtained by  Auch\`ere  (2005).   In addition, the coronal density of the scattering atoms and the local temperature  are  included according to a quiet coronal model (Cranmer et al. 1999).  We perform   a comparison of  the  L$\alpha$ linear polarization  in the  zero-field reference case 
with the amplitude corresponding to the  polarization in the presence of a magnetic field.  In our forward modeling, we adopt a dipolar magnetic distribution as a first step and then we add a magnetic field associated to an equatorial current sheet.

The paper is organized as follows.  We   describe  the theoretical background and formulate  the problem in Sect. 2.   Section 3 deals with the calculations of the Hanle effect without integration over the LOS in order to compare with known results. The generalization of these calculations to  integrate over the LOS  and the discussion of the possibility of obtaining a coronal magnetic field through  polarization measurements are  presented in Sect. 4.     The technique that could be used to measure the scattering polarization of the L$\alpha$ D$_2$ line is given  in Sect. 5; in particular we show how the linear scattering polarization could be measured using a L$\alpha$ disk imager and coronagraph called LYOT (LYman Orbiting Telescope). In Sect. 6 we summarize our conclusions.

 \section{Formulation of the problem}
  \subsection{Hanle effect}
 The term {\it  Hanle effect}  represents the  ways in which the {\it scattering polarization} can be modified by weak magnetic fields. 
The well-known   Zeeman effect and the Hanle effect  are   complementary because they respond to magnetic fields in very different parameter regimes.  The   Zeeman effect   depends on the ratio between the Zeeman splitting and the Doppler line width. The Hanle effect  though depends on the ratio between the Zeeman splitting and the   inverse life time of the  atomic levels involved in the process of the formation of the polarized line.  For  the permitted UV  lines, the Zeeman effect is of limited  interest  for the determination of the magnetic fields in  the quiet corona. This is because   the ratio between the Zeeman splitting and the Doppler   width is small due to the weakness of the   magnetic field  and the high Doppler width in  such  hot coronal plasmas.  On the contrary,  the  measurement and  physical interpretation of  the scattering polarization of the UV  lines are a very efficient diagnostic tool for determining the  coronal magnetic field through  its Hanle effect.  
  \subsection{Atomic linear polarization}
 The possibility of the  creation of a linear polarization by anisotropic scattering can be only explained correctly  in the framework of the  quantum-mechanical scattering theory.  In fact, the intrinsic capacity of a line to be polarized is intimately linked to   subtle quantum behaviors pertaining to  the atomic levels involved in the transition. 
Let us  denote  by $m_J$  the projection of the orbital angular momentum $J$ of the hydrogen atom;   $m_J$  takes the values $-J$, $J+1$,..., $J$.   
 The   term ``atomic linear polarization" in a $J$-level  consists in (e.g. Cohen-Tannoudji \& Kastler 1966, Omont 1977, Sahal-Br\'echot 1977, Blum 1981):
 \begin{description}
\item[--] an unbalance of the populations of the Zeeman sub-levels having different absolute values $|m_J|$
\item[--]   a presence of interferences between  these Zeeman  sub-levels.  
 \end{description}
 This means that  by definition,   only levels having $J>1/2$ can be linearly polarized. 
\subsection{Linear scattering polarization in the L$\alpha$ line}
 The so-called scattering polarization is  simply
  the observational manifestation of the atomic polarization. The {\it Hanle effect} is nothing but a perturbation of the atomic polarization by   a  magnetic field. The Hanle signatures in the spectrum of the linear polarization are a  variation of the polarization degree and a rotation of the polarization plane. These Hanle  signatures can be used to retrieve information on coronal  magnetic fields.  The two components D$_1$ and D$_2$ of the  L$\alpha$ connect the hydrogen ground state $^2S_{J=1/2}$ to the electronic excited states $^2P_{J=1/2}$   and  $^2P_{J=3/2}$, respectively.   The   upper level $^2P_{3/2}$  of the D$_2$  
line     can be  polarized due to the difference of the populations between the Zeeman sub-levels  with $|m_J|=1/2$ and 
$|m_J|=3/2$.  However,    the states $^2S_{1/2}$ and $^2P_{1/2}$  cannot be polarized since  $|m_J|$  is necessarily 1/2 implying that  no difference of   population  inside these states can be generated by anisotropic scattering. Consequently, the D$_1$ line is not linearly polarizable.

It is useful to keep in mind that in the description of the emitting hydrogen atom, we neglect the contribution of the hyperfine structure (HFS).   For instance   if the  HFS is  not neglected, the level  $J = 1/2$ of the ground state $^2S_{1/2}$ is split into hyperfine levels $F = 0$ and $F = 1$ due to coupling with the nuclear spin of the hydrogen $I = 1/2$. The hyperfine level  $F = 1$   can be linearly polarized\footnote{In other words, population  imbalances and quantum interferences between the sub-levels having $|m_F|=1$ and $|m_F|=0$ can be created  due to the scattering of anisotropic light. The same is true for the hyperfine levels of the upper states $^2P_{3/2}$ and $^2P_{1/2}$.}, which means that the D$_1$ line can be polarized and the polarization of the D$_2$ line can be affected.  As previously suggested by Bommier \& Sahal-Br\'echot (1982), we neglect the effect of the HFS in the process of formation of L$\alpha$ line. 

\begin{figure}[htbp] 
\begin{center}
\includegraphics[width=8 cm]{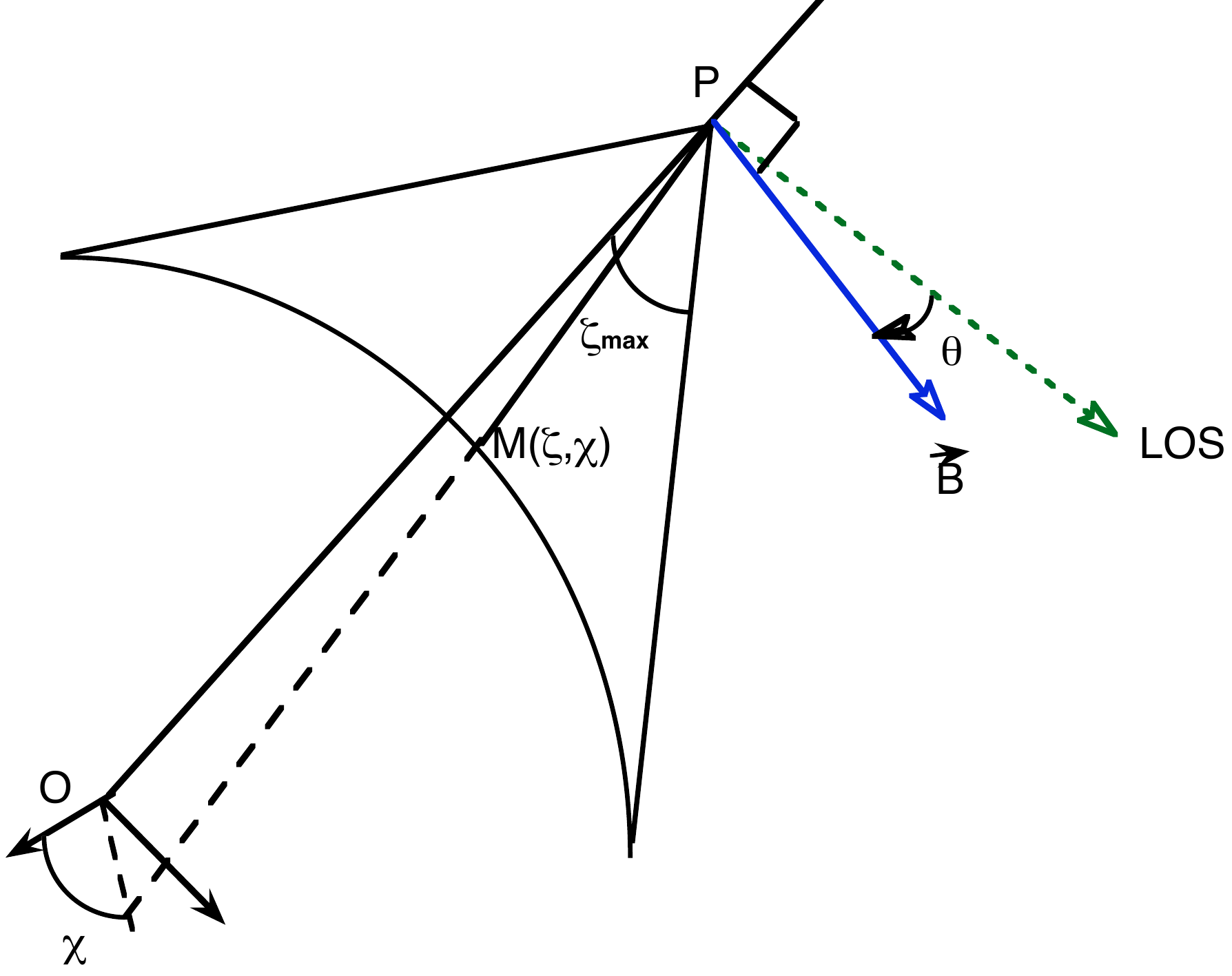}
\end{center}
\caption{ \textsf{ 
Geometry of the scattering of chromospheric L$\alpha$ photons by
residual coronal neutral hydrogen. 
The  anisotropy of the incident 
light is due  to geometrical effects  but also to the 
inhomogeneities of the chromospheric regions.}}
\label{fig1}
\end{figure}

\subsection{Expression of the Stokes parameters} 
The emission of the L$\alpha$  $\lambda$1216 line in the solar corona has been discovered by   Gabriel et al.~(1971). They concluded that   in most coronal structures the  process responsible for the formation  of the L$\alpha$ line is the photo-excitation by underlying radiation.  
 The creation of population  imbalances and the quantum interferences  in the $^2P_{3/2}$  and thus the existence of the scattering polarization in the   D$_2$  L$\alpha$ line   are caused by the  photo-excitation  of coronal neutral hydrogen by anisotropic chromospheric radiation (see Fig \ref{fig1}). 
  
 The components of the incident radiation field at a frequency $\nu_{0}$ are usually denoted by  $\bar{J}_{q}^{k} (\nu_{0}) $  where $k$ is the tensorial order and $q$ represents the coherences in the tensorial basis ($-k \le q \le k$);  the   order  $k$  can be equal to 0  (with $q=0$) or 2 (with $q=0$, $\pm$1, $\pm$2). This radiation field with six components  constitutes a generalization of the  unpolarized   light field where only the quantity $\bar{J}_{0}^{0} (\nu_{0})$ is considered.    In fact, $\bar{J}_{0}^{0} (\nu_{0})$ is proportional to the    intensity of the radiation.  

 If the  incident radiation  is no longer anisotropic, the components $\bar{J}_{q}^{k=2}  (\nu_{0}) $ become  zero, which means that   no linear polarization can be created as a result of scattering processes.  
 Regardless    of the  anisotropy of the incident radiation,  the   radiation component associated     with the circular polarization usually denoted by $\bar{J}_{q}^{k=1}$ is negligible. This   means that no odd order $k$ can  be created inside the scattering hydrogen atom. As a result, the Stokes $V$ of the scattered radiation is zero.

 We denote by $\zeta$ the angle between the direction of the incident light MP  and the local vertical  through 
the scattering center OP. The incident radiation   comes from a chromospheric spherical cap limited by an angle $\zeta_{\textrm{\scriptsize{max}}}$ corresponding to the tangent to the solar limb (see Fig. \ref{fig1}). $\chi$ is the azimuth angle around the normal with respect to an arbitrary reference. Note that  $0 \le$ $\zeta$ $\le$ $\zeta_{\textrm{\scriptsize{max}}}$   and   0 $\le$ $\chi$ $\le$ 2 $\pi$. 
  When the distance from the solar surface increases,  the anisotropy of the light becomes larger and   the polarization degree increases.   The maximum of polarization is reached  when the radiation is purely directive, i.e. the spherical cap is seen by the scattering hydrogen atom as a point.  It is useful to notice that if   the chromosphere is assumed to be uniform the radiation has a cylindrical symmetry around its preferred  direction, implying that the coherence components with $q \ne 0$ are zero. In fact, 
$\bar{J}_{q=\pm1}^{k=2}  (\nu_{0}) $ and $\bar{J}_{q=\pm2}^{k=2}  (\nu_{0}) $ components quantify the breaking of the cylindrical symmetry around the axis of quantification which is here the local vertical.

 In the framework of the two level approximation, where  only the upper level is polarized,     the statistical equilibrium equations  are solved analytically. The upper level density matrix elements are simply proportional to the incident radiation elements $\bar{J}_{q}^{k}$. The emissivity vector  is then expressed  as a function of the incident radiation field. Consequently,  we do not explicitly calculate  the density matrix elements, but instead we determine the incident radiation  tensor  at each scattering position along the line of   sight.  For  an unmagnetized atmosphere,  in an arbitrary reference,  the emissivity vector  can be written as  (e.g. Landi Degl'Innocenti \& Landolfi 2004): 
\begin{eqnarray}  
\epsilon_j (\Omega)  =   n_{\textrm{\scriptsize{H}}}  \frac{h \nu_0 B_{J_lJ_u}}{4 \pi}   \sum_{k,q} W_k(J_l,J_u) \mathcal{T}_q^{k} (j,\Omega) (-1)^q  \bar{J}_{-q}^{k} (\nu_{0})   
 \end{eqnarray}  
where $\Omega$ is  the solid angle giving the direction   of the LOS,  $n_{\textrm{\scriptsize{H}}}$ is the local number density of 
scattering hydrogen atoms, $h$ is 
the Planck constant,   and $B_{J_l,J_u}$ is the Einstein coefficient for absorption. We recall that   $\mathcal{T}_q^{k}
(j,\Omega)$  is the spherical tensor for
polarimetry which contains the angular distribution of the emitted radiation,   and $j$ is the index of the Stokes parameter ($j$ = 0, 1, 2, and 3 for the Stokes $I, Q, U,$ and $V$, respectively). 

 In order to determine the magnetic field one has to include its Hanle effect on the polarization of the L$\alpha$ light, then,  for a given magnetic field vector ${\bf B}$, $\epsilon_j (\Omega) $ becomes   (e.g. Landi Degl'Innocenti \& Landolfi 2004):  
\begin{eqnarray}
\epsilon_j (\Omega,{\bf B})   =  n_{\textrm{\scriptsize{H}}}  \frac{h \nu_0 B_{J_lJ_u}}{4 \pi}    \times
 \end{eqnarray}
 $$
 \sum_{k,q} W_k(J_l,J_u) \mathcal{T}_q^{k} (j,\Omega) (-1)^q  \bar{J}_{-q}^{k} (\nu_{0})  \frac{1}{1+\textrm{i} q H_u} $$
 This expression of $\epsilon_j (\Omega,{\bf B}) $ is correct only in a reference system having the quantization $z$-axis in the magnetic field direction. $H_u$ is the so-called reduced magnetic field strength,  associated to the level $^2P_{3/2}$,   given by:
 \begin{eqnarray}
H_u =    \frac{0.879 \; g_u  \; \textrm{B}}{A_{J_uJ_l}}   
 \end{eqnarray}
 where the   Einstein coefficient for spontaneous emission $A_{J_uJ_l}$ is given in [$10^7$ s$^{-1}$], $g_u=4/3$ is the Land\'e factor of the   level $^2P_{3/2}$ and the magnetic field strength B is given in Gauss.  \mbox{$H_u=1$} corresponds to the magnetic field strength $B=53$ Gauss around which one may expect a noticeable  change in the scattering polarization  of L$\alpha$ with respect to the unmagnetized reference case. The quantity $W_k(J_l,J_u)$ was first introduced by Landi Degl'Innocenti (1984) and depends only on the  quantum numbers of the lower and upper   levels  ($J_l$ and $J_u$)   involved in the transition.  For $k$=2, $W_2(J_l,J_u)$ can be seen as the efficiency  of creation of the linear polarization in the scattering processes. That is why the $W_2(J_l=1/2,J_u=1/2)$ =0 for the D$_1$ line, which is not polarizable, but $W_2(J_l=1/2,J_u=3/2)$=1/2 for the polarizable D$_2$ line.

\section{Hanle effect without integration over the LOS} 
 We developed    a numerical code   allowing for the calculation of the theoretical polarization taking into account the effects of the LOS.  In order to validate the code,   we considered typical  cases of   a horizontal  magnetic field having  different azimuth angles $\theta$  (angles between the magnetic field 
vector and the LOS).    LOS integrations are avoided in order to be able to compare our results with well known Hanle effect results. We   retrieve the Hanle behaviors  typically encountered in the literature, for instance: 
\begin{description}
\item[--]   when the magnetic field is zero or very small or oriented along the symmetry axis of the radiation field, the polarization is not affected  
\item[--]    when the field increases until reaching the critical value corresponding to $H_u=1$, the polarization decreases rapidly. Moreover, for a very large $H_u$ (i.e. very large magnetic field strength) we obtain an asymptotical curve of polarization $p[B \to \infty]$ which depends only on the value of $\theta$ but not on the magnetic field strength. The asymptotic value of $p[B \to \infty]$ divided by $p[B =0]$ equals 1/5 when the distribution of the magnetic field is isotropic\footnote{The case of isotropic field distribution is encountered in the photosphere of the Sun (second solar spectrum) where the magnetic geometries  are unresolved within the spatiotemporal resolution of the current observational capabilities.} and  1/4 when the field has a cylindrical symmetry (i.e. horizontal magnetic field with  random azimuth) 
\item[--]   no rotation of  the plane of the polarization in the case of a highly  symmetric distribution (e.g. isotropic or cylindrical) because the contributions of opposite magnetic polarities tend to cancel out
\item[--]    we find that a meridian magnetic vector (i.e. horizontal with $\theta = \frac{\pi}{2}$) presents a depolarizing effect without rotation of the polarization direction. 
\end{description}
 \section{Hanle effect   integrated over the LOS} 
 The corona being  an optically thin medium for the L$\alpha$ line, it is then necessary to consider the effects of the integration over the   LOS.  We adopt the analytical magnetic field model proposed   by Fong et al. (2002) and Low et al. (2003). It is a sum of two terms: a purely dipolar   term   and a term corresponding to the magnetic field of a current sheet structure. The model is axisymmetric and the prominence is treated as a cold plasma sheet forming a flat ring around the Sun. We take into consideration that the current is in the equator and that it represents a prominence sheet extending from $r=R_{\sun}$  to   $r= \sqrt{4/5} \; R_{\sun}$. 
 In the analytical expression of the magnetic field  the contribution of the current sheet, relative to the dipolar background field  is controlled by a constant ratio $\gamma$ (see Eq. 12 of Low et al. 2003).

  \begin{figure}[htbp] 
\begin{center}
\includegraphics[width=8 cm]{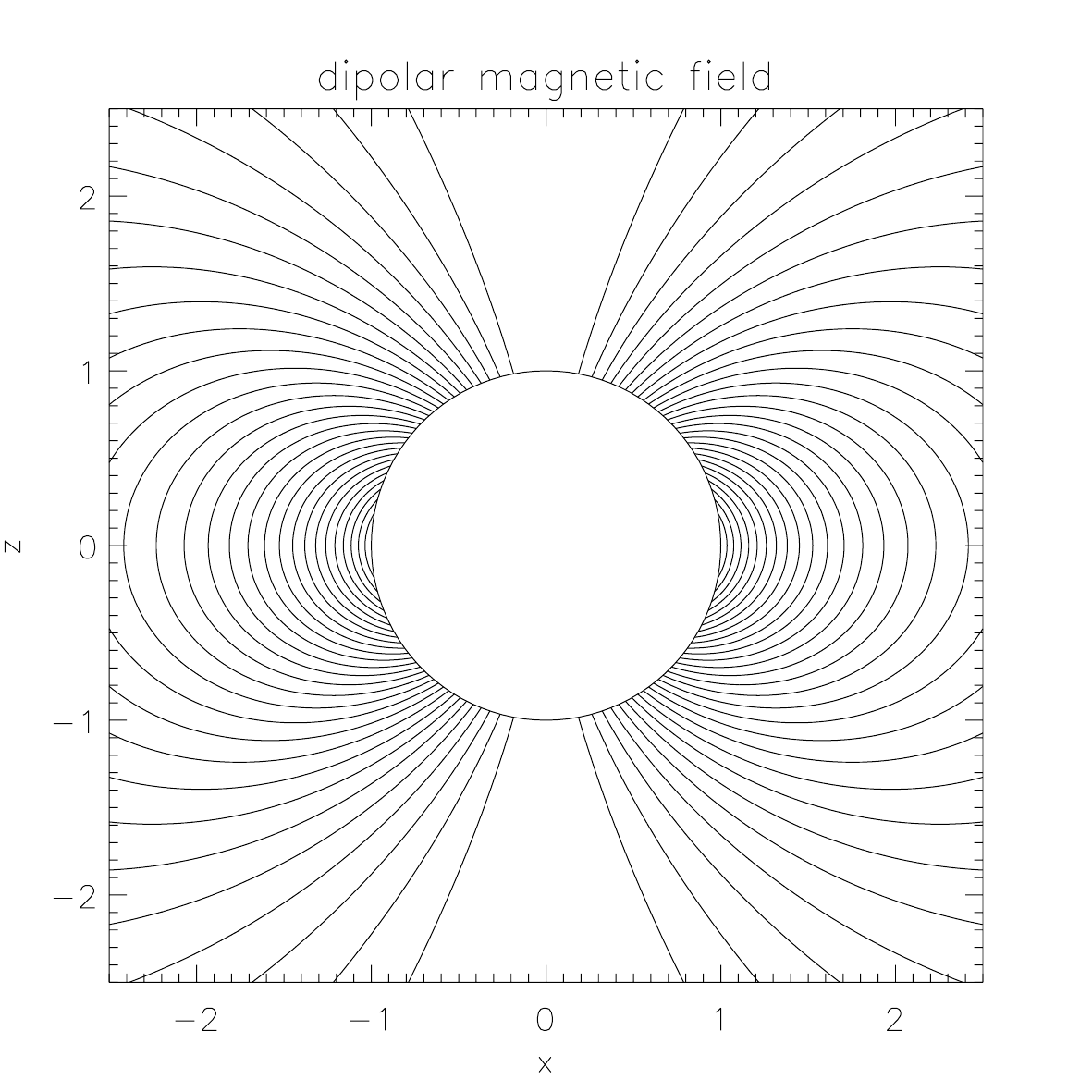}
\end{center}
\caption{ \textsf{  
A purely dipolar magnetic field structure   presented in the plane $(x,z)$ in   units of the solar radius $R_{\sun}$. We use a system of orthogonal Cartesian coordinates $(x, y, z)$ with the origin at  the Sun  center and the   $z$-axis pointing toward the north solar pole.}}
\label{dipolar}
\end{figure}
\subsection{Purely dipolar magnetic field: $\gamma$=0}
\begin{figure}[htbp] 
\begin{center}
\includegraphics[width=8 cm]{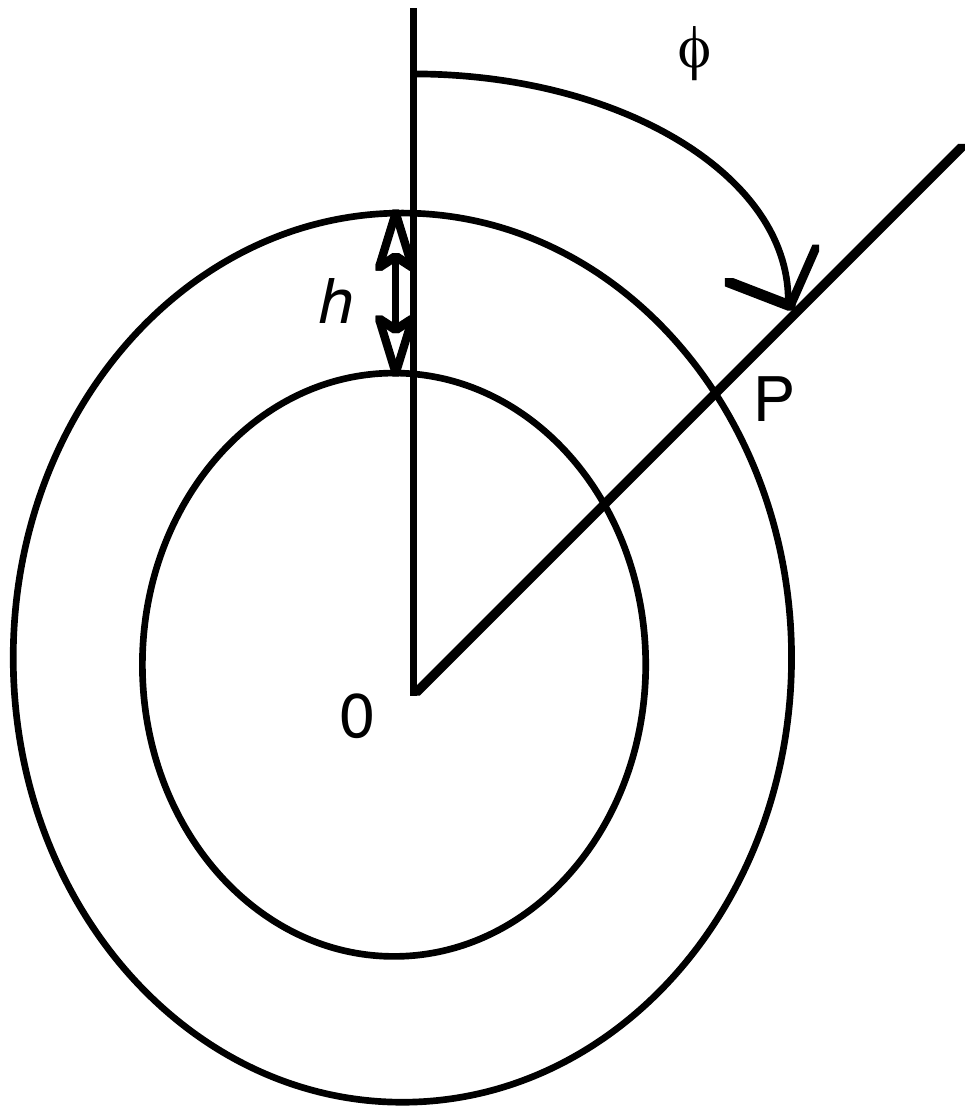}
\end{center}
\caption{ \textsf{ 
 The polarization at each position P of the scattering event   depends on the height over the limb $h$ and the angle $\phi$ between the preferred axis of the radiation and the symmetry axis of the magnetic structure. The LOS is perpendicular to the plane of the figure.}}
\label{HandPHI}
\end{figure}

As a first step, we avoid the effect of the term associated to the current sheet by taking $\gamma$=0. Figure \ref{dipolar} represents the dipolar term  of the magnetic field. This configuration represents a typical coronal magnetic field of 15 to 20  Gauss close to the base of the corona. The Hanle 
effect of the dipolar magnetic field depends on the angle $\phi$ between the axis of symmetry of the incident light\footnote{Rigorously speaking, this is only  the axis of symmetry  of the spherical cap where the chromospheric radiation is uniform. The fact that the incident radiation is  inhomogeneous implies that this  symmetry  around the preferred axis of radiation is broken. } and the axis of symmetry of the magnetic structure. It also depends  on the height above the solar surface $h$ mainly because the magnetic field strength decreases and the anisotropy of the incident light increases. The parameters $\phi$ and $h$ are  represented in   Fig. \ref{HandPHI}.

\begin{figure}[htbp] 
\begin{center}
\includegraphics[width=8 cm]{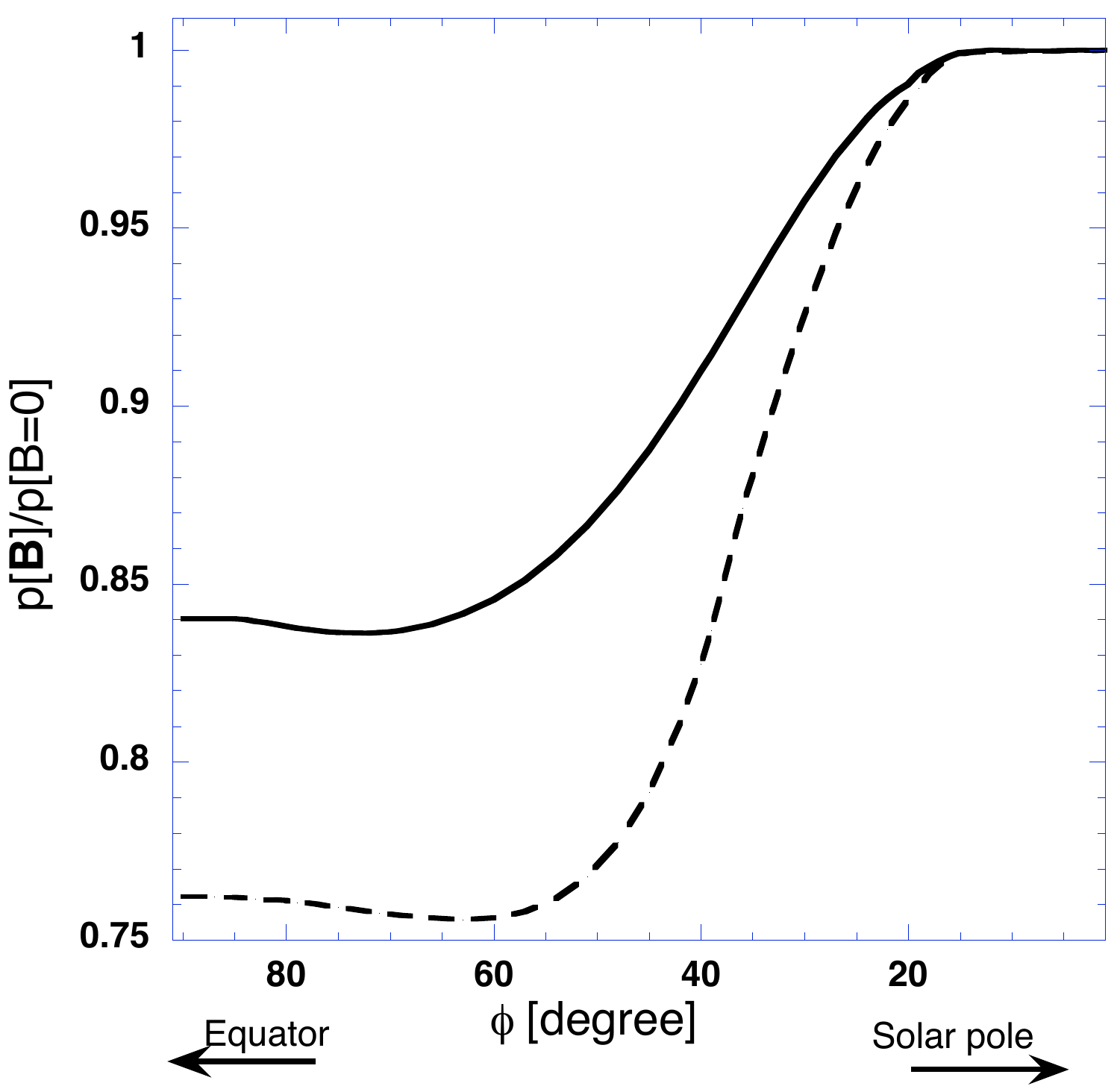}
\end{center}
\caption{ \textsf{ 
 Linear polarization degree obtained for a dipolar magnetic field divided by zero-field polarization  versus the angle $\phi$.   Full lines   represent   the polarization at $h$=0.5 R$_{\sun}$ and   dashed lines represent  the polarization at  $h$=0.3 R$_{\sun}$.   }}
\label{polar}
\end{figure}

\begin{figure}[htbp] 
\begin{center}
\includegraphics[width=8 cm]{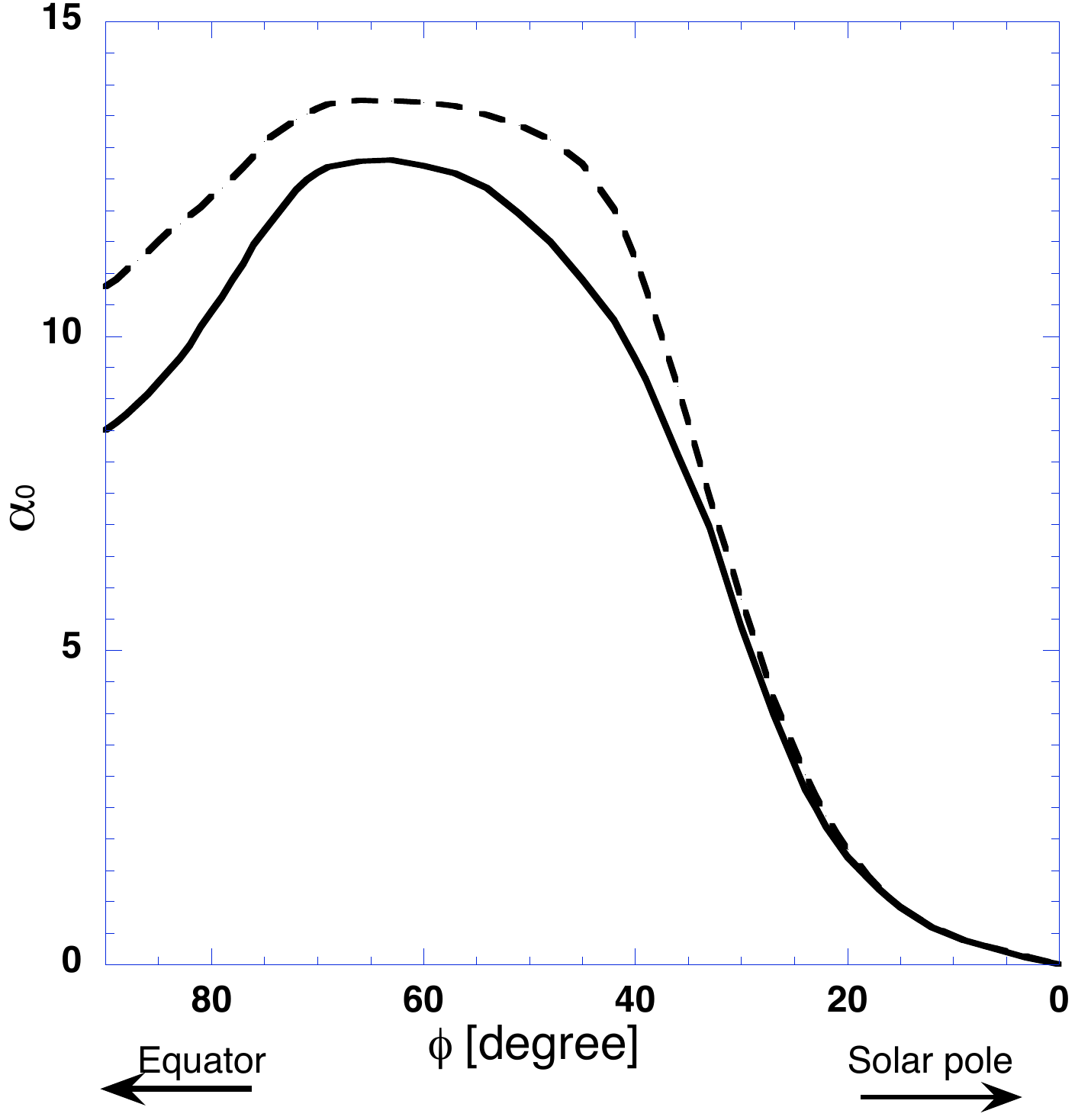}
\end{center}
\caption{ \textsf{ 
Rotation angle obtained after integration over the LOS versus the   inclination $\phi$    at $h$=0.3 R$_{\sun}$ and $h$=0.5 R$_{\sun}$.  Full lines   represent       the rotation at $h$=0.5 R$_{\sun}$ and   dashed lines represent the rotation   at  $h$=0.3 R$_{\sun}$. }}
\label{rota}
\end{figure}

In theory, the expression of the emissivity vector is valid regardless of the location of the scattering atom. However, the integration over the LOS must take into account the inhomogeneities of the solar conditions like the variation of the hydrogen density, the temperature, the magnetic field and the variation of the incident radiation field.  The density of neutral hydrogen and the temperature  are assumed to be a function of the radial distance $r$ and the latitude (see Cranmer et al. 1999 for details). In order to model the  inhomogeneities of the chromospheric intensity, we  use  the Carrington maps of the L$\alpha$ chromospheric line built by Auch\`ere (2005). In the optically thin limit,    the integrated Stokes parameters   of the  scattered radiation reduce  to a volume integration  over the LOS:
\begin{eqnarray} 
\mathcal{E}_j (\Omega) &= &  \int_{\textrm{\scriptsize{LOS}}} \epsilon_j (\Omega) dl  
\end{eqnarray}
then the polarization degree is 
\begin{eqnarray} 
p &= &  \frac{\sqrt{\mathcal{E}_{1}^2+\mathcal{E}_{2}^2}}{\mathcal{E}_0}
\end{eqnarray}
and the rotation of the direction of the polarization $\alpha_0$ is given by:
\begin{eqnarray} 
tg(2 \alpha_0) &= & \frac{\mathcal{E}_2}{\mathcal{E}_1} 
\end{eqnarray}
Figure \ref{polar} shows the variation of the linear polarization with the inclination $\phi$    for two  different   magnetic structures corresponding  to two  heights above the solar surface. Note that the ratio $p[B]/p[B=0]$ obtained for  the height $h$=0.3 R$_{\sun}$ is   smaller  than the one corresponding to $h$=0.5 R$_{\sun}$ since the magnetic field decreases with $h$. Furthermore, as shown in Fig. \ref{rota},  a notable Hanle rotation of about $10^o$ is obtained  for  $h$=0.3 R$_{\sun}$ and for $h$=0.5 R$_{\sun}$. Both Hanle signatures on the L$\alpha$   line, i.e.  depolarization and rotation (see Figs. \ref{polar} and   \ref{rota}), are clearly sizable  for $\phi >$ $40^o$.  This important result   suggests  that  in order to measure a dipolar magnetic field by its Hanle effect one should observe regions rather far away from the pole.

\begin{figure}[htbp] 
\begin{center}
\includegraphics[width=6 cm]{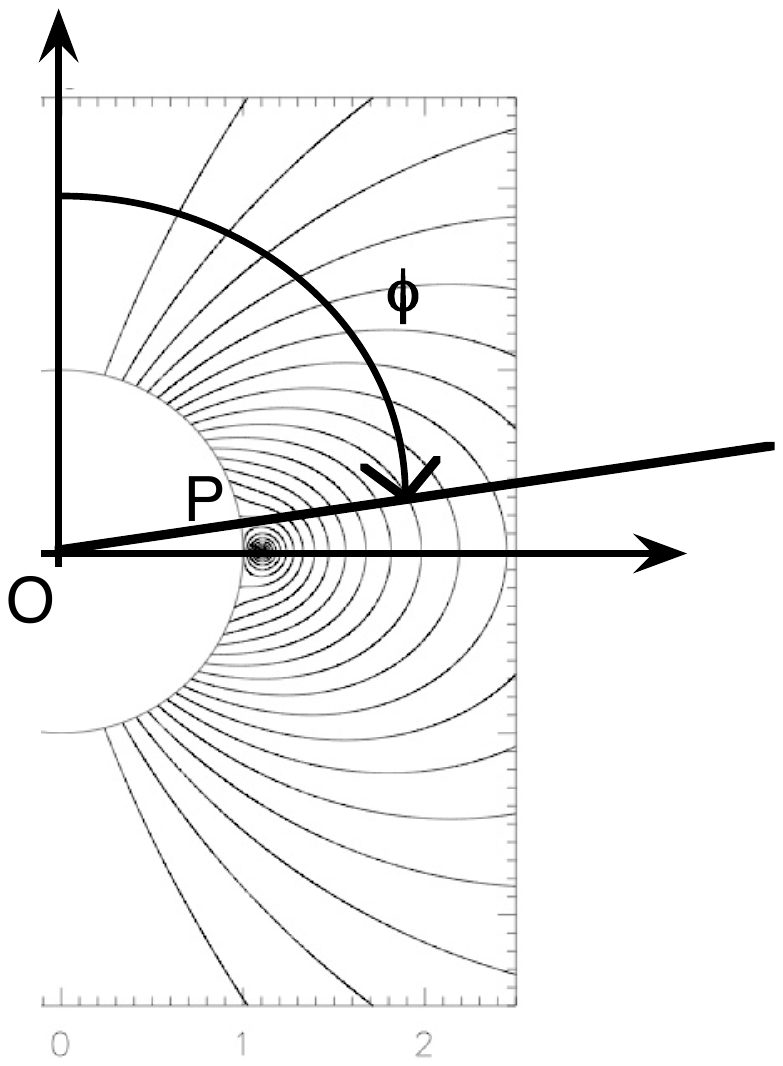}
\end{center}
\caption{ \textsf{ 
Perturbation of the lines of the dipolar magnetic field due to an equatorial    current sheet.  The calculations of the polarization   given in Fig. \ref{polarALL} are obtained for $\phi=$$80^o$. }}
\label{MagCS}
\end{figure}

  \begin{figure}[htbp] 
\begin{center}
\includegraphics[width=6 cm]{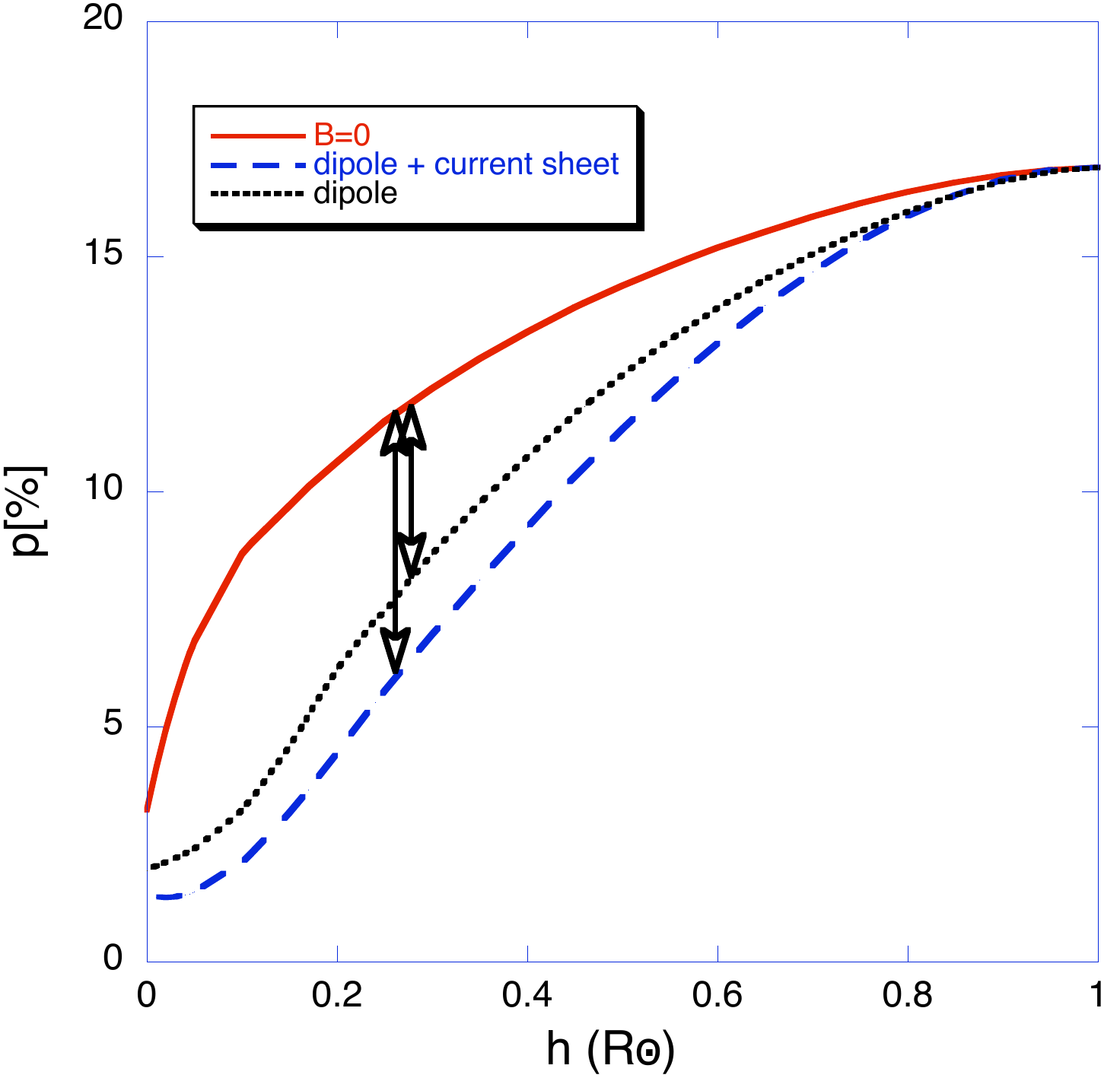}
\end{center}
\caption{ \textsf{ 
Linear polarization degree versus the  height from the solar surface $h$.   We put together the results obtained in the zero-field case and these obtained for (1) a  purely dipolar magnetic  field  (2)
the sum of a   dipolar   field  and a non-dipolar field associated to a current sheet  with $\gamma=$ 0.25.   
  }}
\label{polarALL}
\end{figure}

\subsection{Perturbed dipolar  field: $\gamma \ne$ 0}
To the dipolar part of the magnetic field we now  add the   contribution resulting from an equatorial current sheet.    We adopt a ratio $\gamma$=0.25 between the current sheet   and the dipolar background field.    In order to highlight the Hanle effect  of the equatorial current sheet,  we calculate the degree   of polarization in a position  located 
at $\phi$=$80^o$, and we vary the height above the limb $h$ (see Fig. \ref{MagCS}).   Figure \ref{polarALL} shows    the difference  between the linear polarization in the zero-magnetic field case and 
the one in the presence of the magnetic field, $\Delta p$=$|p_{B=0} - p_{B \ne 0}|$. A polarimetric sensitivity smaller than $\Delta p$  is needed in order 
to apply the Hanle effect as a technique of    magnetic field investigations.  Our results show that   $\Delta p$ $\sim$  5 \%, i.e.  well   within 
the  typical measurement sensitivities  of a new  generation of instruments  such as LYOT (see Sect. 5).
We  point out that by using the  UV  SUMER  spectrometer aboard SoHO, Raouafi et al. (1999) measured the linear polarization of the O {\sc vi} $\lambda$1032 line with a  polarimetric precision equal  to \mbox{1.7 \%}. We note in passing that   such an accuracy is reached although    SUMER  was not initially designed to 
measure the polarization. 
\section{ Measurement  of the  linear polarization degree and its direction } 
\subsection{ Principle of the measurement}
Raouafi et al. (1999)  used the rotation of the SUMER spectrometer to measure the linear polarization of the D$_2$ component of the 
O {\sc vi}  $\lambda$1032 line. They extracted the polarization of the D$_2$ line from a  ratio  of the intensities of the non-polarizable D$_1$ line  and of the D$_2$ line (see the  Fig. 3 of Raouafi et al.  1999).
This technique was possible because the wavelengths of the two components D$_1$  and  D$_2$ are sufficiently different to  be resolved (1031.93 \AA\ for the D$_2$ line and 1037.62 \AA\ for the D$_1$). However, 
 the wavelengths of the   D$_1$ and D$_2$ lines of the L$\alpha$    line  cannot be resolved since they are  very close: in the vacuum  $\lambda$(D$_1$)= 1215.668 \AA\  and  $\lambda$(D$_2$)= 1215.674 \AA. As a result, the technique presented by Fig. 3 of Raouafi et al.  (1999)  cannot be applied to measure the linear polarization of the    D$_2$ line of L$\alpha$.

Using the so-called Poincar\'e representation\footnote{A suitable graphical representation of polarized light  conceived by Henri Poincar\'e in 1892.},  one can demonstrate that the intensity observed when the instrument   is placed at an arbitrary position referred by an angle $\beta$ around the LOS, is
\begin{eqnarray} 
I(\beta) = \frac{1}{2} (Q \cos 2\beta + U \sin 2\beta + I)
\label{eq3}
\end{eqnarray}
In this expression the Stokes $V$ is assumed to be zero. The quantity $I$ denotes the unpolarized part of the intensity of the D$_1$ and D$_2$ L$\alpha$ lines. 
$I(\beta)$ represents the ``real" (polarized and unpolarized) observed intensity of the two resonance  lines. In addition, taking into account that $\alpha_0$ corresponds to the direction of linear polarization (i.e. privileged direction of the electric field), Eq. (\ref{eq3}) becomes
\begin{eqnarray} 
I(\beta) = \frac{1}{2} (\sqrt{Q^2 + U^2} \cos 2(\beta-\alpha_0)+ I)
\label{eq4}
\end{eqnarray}
Note that  $\cos 2\alpha_0  = \frac{Q}{\sqrt{Q^2 + U^2}}$ and $\sin 2\alpha_0  = \frac{U}{\sqrt{Q^2 + U^2}}$. 

On the other hand, generally speaking, the linear polarization is defined as
\begin{eqnarray} 
p = \frac{I_{max}-I_{min}}{I_{max}+I_{min}} 
\label{eq5}
\end{eqnarray}
 where $I_{max}$ and $I_{min}$ are the maximum and minimum intensities.  Using Eqs. (\ref{eq4}) and (\ref{eq5}), one finds that
  \begin{eqnarray} 
p = \frac{\sqrt{Q^2 + U^2}}{I}
\end{eqnarray}
and
\begin{eqnarray} 
\frac{I(\beta)}{I}  = \frac{1}{2} (p\cos 2(\beta-\alpha_0)+1)
\label{eq7}
\end{eqnarray}
In Eq. (\ref{eq7}) we have three unknowns: $I$, $p$, and $\alpha_0$. Then, theoretically,  the linear polarization state is fully obtained through   only three measurements of the $I(\beta)$ which corresponds to three rotations of the polarizer-spectrometer.   One takes for example $\beta=0,  \frac{\pi}{4},$ and $ \frac{\pi}{2}$, therefore:
\begin{eqnarray}
\tan 2\alpha_0  &= &  \frac{2 I(\frac{\pi}{4}) -I}{I(0) - I(\frac{\pi}{2})}   \\ 
p  &= &   \frac{I(0) - I(\frac{\pi}{2})}{\cos 2\alpha_0 \times I} = \frac{2 I(\frac{\pi}{4})-I}{\sin 2\alpha_0 \times I}   \nonumber 
\end{eqnarray}
where   the intensity of the unpolarized light is given by:  
\begin{eqnarray}
I &= & I (0) + I (\frac{\pi}{2})  
\end{eqnarray}
 Obviously, more than three  measurements of $I(\beta)$ are welcome  in order to increase the accuracy.
\subsection{ The LYOT  project}
 The LYOT  project   is   a L$\alpha$  coronagraph  combined with  a L$\alpha$ disk imager   (see Vial et al.~2002,  Millard et al.~ 2006, and  Vial et al.~2008).   In addition,  it is planned  to implement a simple polarizer system. The polarizing measurements will be performed by rotating the polarizer or the whole   instrument  to  obtain     the intensity of the L$\alpha$ light at different $\beta$ angles  (previous section).
The choice of the L$\alpha$ line is well justified 
 by its  sensitivity   to the coronal magnetic field (as   demonstrated  in this paper) 
 and by the fact that    in the corona   the L$\alpha$ emission is very intense.    In fact, a high signal to noise ratio is needed since in the very low corona the anisotropy of the light is small, which in turn means that the polarization degree is small (smaller than  5 \%, see Fig. \ref{polarALL}). We note that no coronagraph observing as low as 1.15 R$_{\sun}$ is envisaged beyond 2012 except for LYOT  images  which should be obtained with an excellent  signal to noise ratio.

 \section{Conclusion}
Measurement and   interpretation of the scattering polarization of    UV coronal lines provide a largely unexplored diagnostic of the coronal magnetic field. The greatest difficulty facing the UV  coronal spectropolarimetry is that the polarization measurements    integrate radiation along the LOS over   structures with different  properties   but also that the observations of these lines are impossible  from   ground-based telescopes;   they can only be observed   with the help of  high-sensitivity    instruments  flown on space missions.

We have performed   a forward  modeling of the coronal Hanle effect on the polarization of the L$\alpha$ line generated by anisotropic scattering of chromospheric light.
The main feature of this modeling consists in integrating the effect of the LOS. We show that the information about the coronal magnetic field is not lost through LOS integration.   To confirm these results,   we plan to work with   different families of maps of magnetic fields and to add small scale magnetic perturbations. For instance, one can think of a set of active loops whose field determination could be compared with field extrapolations. One should however  keep in mind that (1) a realistic thermodynamic model  is required in order to integrate along the LOS and that (2)  our modeling is limited to the case of optically thin plasmas in the L$\alpha$ line.

Finally, we notice that it is suitable to combine measurements in L$\alpha$  with measurements  in polarized  lines like  the  Fe {\sc xiv}    $\lambda$5303  which have a different  sensitivity to the magnetic field. An analysis of combined measurements should give  more  information  data to constrain  the  magnetic field topology and strength   (an example of the Hanle effect in a multi-line approach is given in Landi Degl'Innocenti 1982). It is also of interest to remark   that the ground level  $^2S_{J}$ of L$\alpha$ is non polarizable by radiation anisotropy, but that this is no longer true  in the  presence of hyperfine structure and if the depolarizing effect of the isotropic collisions is negligible.  Because the sensitivity to the Hanle effect depends on the level life-time,  the hyperfine polarization of   such a long-lived level is much more vulnerable to  very weak magnetic fields  than the short-lived upper levels   $^2P$.   Consequently, one could distinguish   a  very small perturbation of the magnetic field (smaller than 1 Gauss) which corresponds for instance to a current sheet with a very small 
$\gamma$ and a background field of the order of 10 Gauss or larger.  In particular, this could be the key  to distinguish   potential magnetic field structures from non potential ones.  


\end{document}